\documentstyle[12pt,epsfig]{article}
\input axodraw.sty
\def\lsim{\:\raisebox{-0.5ex}{$\stackrel{\textstyle<}{\sim}$}\:}
\def\gsim{\:\raisebox{-0.5ex}{$\stackrel{\textstyle>}{\sim}$}\:}

\begin{document}
\begin{flushright}
BI-TP 97/25\\[1.7ex]
{\tt hep-ph/9707447} \\
\end{flushright}

\vskip 45pt
\begin{center}
{\Large \bf Generalized Vector Dominance and low $x$ \\[1ex]
inelastic electron-proton scattering}$^{\displaystyle \ast}$

\vspace{11mm}

{\large D. Schildknecht and H. Spiesberger}\\[2ex]
{\em Fakult\"at f\"ur Physik, Universit\"at Bielefeld, 
D-33501 Bielefeld}\\[2ex]

\vspace{150pt}

{\bf ABSTRACT}
\end{center}
\begin{quotation}
The HERA data on inelastic lepton proton scattering for values of the 
Bjorken scaling variable $x \lsim 0.05$ are confronted with predictions
based on Generalized Vector Dominance. Good agreement between theory
and experiment is found over the full kinematic range of the squared
four-momentum-transfer, $Q^2$, from $Q^2 = 0$ (photoproduction) to
$Q^2 \lsim 350$ GeV$^2$.
\end{quotation}

\vspace*{\fill}
\footnoterule
{\footnotesize
  \noindent ${}^{\displaystyle \ast}$ Supported by the Bundesministerium
  f\"ur Bildung und Forschung, Bonn, Germany, Contract 05 7BI92P (9) and
  the EC-network contract CHRX-CT94-0579.}

\newpage

In the 1972 paper by Sakurai and one of the present authors
\cite{sakurai} it was conjectured that the large cross section then
observed \cite{bloom} in deep inelastic electron scattering at large
values of $\omega^\prime$ (i.e., for small values of $x \simeq Q^2/W^2$
in present-day notation) was caused by contributions from vector states
more massive than $\rho^0, \omega, \phi$ to the imaginary part of the
virtual Compton forward scattering amplitude (compare Fig.\ 1).  The
dominant role of $\rho^0 , \omega , \phi$, which saturate the imaginary
part of the forward Compton amplitude in photoproduction at the level of
78 \% \cite{sakurai}, with increasing spacelike $Q^2$ was expected to be
rapidly taken over by more massive states coupled to the photon. It was
shown, that the deep inelastic electron scattering data available at the
time in the region of $x \lsim 0.15$ were in accord with this hypothesis
of Generalized Vector Dominance. If this hypothesis were viable, it was
argued, that in generalization of the role of the low-lying vector
mesons $\rho^0 , \omega , \phi$, in (real) photoproduction, high-mass
states should be diffractively produced in deep inelastic scattering at
small $x$. Accordingly it was suggested to look for the production of
such states and to compare their properties with the ones of the final
state in $e^+e^-$ annihilation experiments still in the state of
planning in 1972. Moreover, as a further test of this picture for deep
inelastic scattering at small $x$, shadowing in the scattering from
complex nuclei, observed in photoproduction at the time, was predicted
\cite{schild1} to persist at small $x$ in deep inelastic scattering.

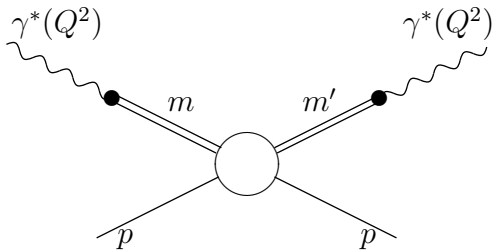
\begin{figure}[htbp]
\begin{picture}(260,120)(-110,0)
\Photon(20,95)(60,75){2}{4}
\put(60,75){\circle*{6}}
\Line(61,76)(101,56)
\Line(59,74)(99,54)
\put(111,50){\circle{25}}
\Line(121,56)(161,76)
\Line(122,54)(162,74)
\put(161,75){\circle*{6}}
\Photon(161,74)(201,94){2}{4}
\Line(54,22)(100,45)
\Line(121,45)(167,22)
\put(22,100){$\gamma^*(Q^2)$}
\put(170,100){$\gamma^*(Q^2)$}
\put(62,20){$p$}
\put(154,20){$p$}
\put(81,70){$m$}
\put(132,70){$m'$}
\end{picture}
\caption{\it The virtual Compton forward scattering
  amplitude in Generalized Vector Dominance. The spacelike photon $(Q^2
  \geq 0)$ virtually dissociates into $q \bar q$ states of masses $m$,
  $m'$ which undergo forward scattering on the target proton.}
\label{fig1}
\end{figure}

Shadowing in electron scattering, after many years of confusion, was
established \cite{emc} in recent years in semi-quantitative agreement
with expectation \cite{bilchak}. Moreover, large-rapidity gap events
with typically diffractive features and masses much higher than the
masses of $\rho^0 , \omega , \phi$ have been observed \cite{h1zeus} at
HERA. Encouraged by these experimental results, which are in qualitative
accord with expectations based on Generalized Vector Dominance, in the
present paper we examine the question in how far predictions based on
Generalized Vector Dominance are in quantitative agreement with the
results on the proton structure function at low $x$. 

More specifically, it is the purpose of the present note to show that
the simple 1972 ansatz \cite{sakurai}, appropriately generalized to take
care of the rise of (hadronic) cross sections with energy, not yet known
in 1972, yields predictions for the low $x$ proton structure function
which are in good agreement with the experimental data from HERA
\cite{f2data}.

The starting point of Generalized Vector Dominance for a theoretical
description of the transverse photon absorption cross section $\sigma_T
(W^2, Q^2)$ at small values of $x$ is the mass dispersion relation
\cite{sakurai}
\begin{equation}
\sigma_T (W^2, Q^2) = \int dm^2 \int dm^{\prime 2} {{\tilde \rho_T (W^2;
m^2, m^{\prime 2}) m^2 m^{\prime 2}} \over {(m^2 + Q^2) (m^{\prime 2} +
Q^2)}},
\end{equation} 
where the double spectral weight function $\tilde \rho_T (W^2; m^2,
m^{\prime 2})$ is proportional to the imaginary part of the forward
amplitude for $V + p \to V^\prime + p$ where $V$ and $V^\prime$ are
states of masses $m$ and $m^\prime$ coupled to the photon. In (1), we
have adopted the usual conventions, in which the total center-of-mass
energy of the virtual-photon-proton system, $\gamma^*p$, is denoted by
$W$, the squared four-momentum transfer to the proton by $Q^2$ and the
Bjorken scaling variable by $x = Q^2/(W^2 - M^2_p + Q^2) \simeq Q^2/W^2$
for $W^2 \gg Q^2$.  If the diagonal approximation
\begin{equation}
\tilde \rho_T (W^2; m^2, m^{\prime 2}) = \rho_T (W^2, m^2) \delta (m^2
- m^{\prime 2})
\end{equation} 
is adopted, (1) takes the particularly simple form
\begin{equation}
\sigma_T (W^2, Q^2) = \int_{m^2_0} dm^2 {{\rho_T (W^2, m^2) m^4} \over
{(m^2 + Q^2)^2}},
\end{equation} 
where the threshold mass, $m_0$, is to be identified with the mass scale
at which the cross section for the process $e^+ e^- \to$ hadrons starts
to become appreciable.  As depicted in Fig.\ 1, the spectral weight
function $\rho_T (W^2, m^2)$ in (3) is proportional to
\begin{itemize}
\item[i)] the transition strength of a timelike photon to the hadronic
  state of mass $m$ as observed in $e^+e^-$ annihilation at the energy
  $\sqrt s \equiv m$, and
\item[ii)] the imaginary part of the forward scattering amplitude of
  this state of mass $m$ on the nucleon.
\end{itemize}
For real photons, $Q^2 = 0$, the cross section $\sigma_T (W^2, Q^2 = 0)
\equiv \sigma_{\gamma p}$ in the representation (3) is almost saturated
(up to $\simeq 78 \%$ \cite{sakurai}) by contributions from the discrete
vector meson states, $\rho^0 , \omega$ and $\phi$. For spacelike $Q^2$,
the contribution of $\rho^0 , \omega , \phi$ becomes rapidly
unimportant, and their role, according to Generalized Vector Dominance,
at small $x$ is taken over by a sum of those more massive states which
are produced in $e^+e^-$ annihilation experiments beyond $\rho^0,
\omega, \phi$ at the energy of $\sqrt s \equiv m$.

Before elaborating on the ansatz (3) let us also quote its
generalization \cite{sakurai} to the longitudinal photon absorption
cross section, $\sigma_L$.  A priori, the ratio of high-mass
longitudinal-to-transverse forward scattering amplitudes for hadronic
states of mass $m$ evolving from $q \bar q$ vector states is unknown and
has to be left open by introducing a parameter $\xi$ for this ratio.
Moreover, generalizing $\rho^0$ dominance, where the coupling of the
$\rho^0$ to a conserved source implies a factor $Q^2/m^2_\rho$ for
longitudinal photons \cite{fraas1}, a factor $Q^2/m^2$ is introduced in
the ansatz for $\sigma_L$. It assures the required vanishing of
$\sigma_L$ for $Q^2 \to 0$. Accordingly,
\begin{equation}
\sigma_L (W^2, Q^2) = \int_{m^2_0} dm^2 {{\xi \rho_T (W^2, m^2) m^4}\over
{(m^2 + Q^2)^2}} {{Q^2} \over {m^2}}.
\end{equation} 
Let us note that, first of all, the diagonal approximation, and secondly
the equality of the spectral weight functions in (3) and (4) apart from
the factors $Q^2/m^2$, and $\xi$, as well as the constancy of the factor
$\xi$, constitute fairly drastic assumptions in view of the fact that
with increasing $Q^2$ higher and higher masses, $m$, will dominate the
integrals in (3) and (4).

The underlying assumptions can be empirically tested by analyzing the
observed diffractive production (large rapidity gap events) as a
function of $Q^2$ and $m^2$ at momentum transfer $t \to 0$.  We note
that $\rho^0$ production already, i.e. the contribution due to $m =
m_\rho$ in (3) and (4), at large $Q^2$ does not exactly follow
\cite{rho} the $\rho^0$-dominance form of the diagonal approximation.
This in itself does not contradict the approximate validity of (3) and
(4), as the $\rho^0$ contribution becomes rapidly unimportant as soon as
$Q^2 \gg m^2_\rho$, but it may indicate that the ansatz (3) and (4)
needs refinements in the future, e.g. by allowing for a mass dependence
of $\xi$, or by modifying the factor $Q^2/m^2$ in the region of $Q^2 \gg
m^2$, or by allowing for off-diagonal terms \cite{fraas2} according to
(1). As for the time being, we will see that the gross features of the
HERA data on the sum of $\sigma_T$ and $\sigma_L$, i.e. on $F_2$, will
be adequately represented by the very simple forms (3) and (4), once
$\rho_T$ will have been specified.

In the present paper, we concentrate on deep inelastic scattering in the
high-energy region of $W \gsim 30$ GeV with $Q^2$ between zero and an
upper limit determined by the restriction to sufficiently low values of
$x \lsim 0.05$, where the dynamical assumptions of the present work are
expected to hold.  In the energy range of $W \gsim 30$ GeV hadronic
cross sections as well as photoproduction ($Q^2 = 0$) rise with
increasing energy. Accordingly, we adopt a logarithmic rise\footnote{A
  power behavior \cite{dola1}, $\rho_T \propto (W^2)^{\alpha}$, would be
  possible as well. However, it turns out that when fitting the
  resulting expression for $F_2$ to the HERA data, only data in a
  restricted range of $x$ and $Q^2$ values can be described.} of the
imaginary part of the forward scattering amplitude of the state of mass
$m$ on the proton in addition to a $1/m^2$ fall from dimensional
analysis\footnote{Bjorken \cite{bjorken} conjectures that this $1/m^2$
  fall may be the consequence of the fact that only those $q\bar{q}$
  configurations interact with the proton which are properly aligned in
  the direction of the incident virtual photon in the photon-proton rest
  frame.  Alternatively, off-diagonal transitions implying an effective
  $1/m^2$ behavior have been suggested \cite{fraas2}.}, and an
additional $1/m^2$ decrease from $e^+e^-$ annihilation due to the
couplings of the photon to the initial and final states in Fig.\ 1, i.e.
\begin{equation}
\rho_T (W^2, m^2) = N {{\ln (W^2/am^2)}\over {m^4}}.
\end{equation} 
The parameter $N$ contains the normalization of the cross section for
$e^+e^- \to$ hadrons and determines the overall normalization of the
cross section for the scattering of the state of mass $m$ on the proton.
The parameter $a$ sets the scale for the logarithmic $W$ dependence of
the forward scattering amplitude for the scattering of the state of mass
$m$. Obviously, the scale of the logarithmic $W$ dependence need not
coincide with the mass $m$ of the hadronic vector state being scattered.
We will assume that $a$ is constant, independent of $m$ as well as
independent of the quark flavor in the state of mass $m$.

The transverse photon absorption cross section, $\sigma_T$, then becomes
\begin{equation}
\sigma_T (W^2, Q^2) = N \int^\infty_{m^2_0} dm^2 {{\ln (W^2/am^2)} \over
{(m^2 + Q^2)^2}}.
\end{equation} 
The ansatz (6) again contains a fairly crude approximation insofar as we
do not separate the contributions due to charm, i.e.  $c \bar c$ states
with a higher threshold of $\overline{m}^2_0 \simeq (3\,{\rm GeV})^2$,
from the contributions due to $q \bar q$ configurations of light quarks,
``dual'' to $\rho^0, \omega, \phi$, with thresholds $m_0^2$ of the order
of $m^2_\rho$ and $m^2_\phi$.  From (6) we obtain
\begin{equation}
\sigma_T (W^2, Q^2) = N \left[{1 \over {Q^2 + m^2_0}} \ln {{W^2} \over 
{am^2_0}}
- {1 \over {Q^2}} \ln \left(1 + {{Q^2} \over {m^2_0}}\right)\right].
\end{equation} 
We also note the limits of photoproduction, $Q^2 \to 0$, in (7)
\begin{equation}
\sigma_T (W^2, Q^2 \to 0) = \sigma_{\gamma p} = {N \over {m^2_0}}
\left( \ln {{W^2} \over {am^2_0}} -1 \right),
\end{equation} 
and of deep inelastic scattering, $Q^2 \gg m^2_0$,
\begin{equation}
\sigma_T (W^2, Q^2 \gg m^2_0) \simeq {N \over {Q^2}} \ln {{W^2} \over
{a Q^2}}.
\end{equation} 
According to (8), the normalization of the experimental photoproduction
cross section determines the ratio $N / m^2_0$, while $a\cdot m_0^2$ is
determined by the scale of its energy dependence. The threshold mass
$m_0$ being essentially fixed by $e^+e^-$ annihilation into hadrons, the
parameter $a$ describes the "hadronlike`` energy dependence of
photoproduction. 

The longitudinal photon absorption cross section, $\sigma_L (W^2, Q^2)$,
according to (4) and (5) becomes
\begin{equation}
\begin{array}{lcc}
\sigma_L (W^2, Q^2) &=& \displaystyle
N \xi \Biggl[ \left({1 \over {Q^2}} \ln \left(1 + {{Q^2} 
\over {m^2_0}}\right) - {1 \over {Q^2 + m^2_0}} \right)
\ln {{W^2} \over {am^2_0}}\\[2ex]&&\displaystyle
+ {1 \over {Q^2}} \left(\ln \left(1 + {{Q^2} \over {m^2_0}}\right) + 
{\rm Li}_2 \left(- {{Q^2}
\over {m^2_0}}\right) \right)\Biggr].
\end{array}
\end{equation} 
${\rm Li}_2$ denotes the dilogarithm defined by ${\rm Li}_2 (z) = -
\int^z_0 dt \ln (1-t)/t$. The longitudinal cross section (10) 
vanishes as $Q^2 \ln Q^2$ for $Q^2 \to 0$, while for $Q^2 \gg m^2_0$,
\begin{equation}
\begin{array}{lcc}
\sigma_L (W^2, Q^2 \gg m^2_0) &=& \displaystyle
{{N \xi} \over {Q^2}} \Biggl\{\left( \ln {{Q^2} \over
{m^2_0}} - 1 \right) \ln {{W^2} \over {aQ^2}}\\[2ex]&&\displaystyle
- {1\over 2} \ln^2 {{Q^2} \over {m^2_0}} - {{\pi^2} \over 6}\Biggr\}, 
\end{array}
\end{equation} 
where the asymptotic formula $\lim_{z \to \infty} {\rm Li}_2 (-z) = -
{1 \over 2} \ln^2 z - {{\pi^2} \over 6} + O(z)$ was used.

Before turning to the comparison with experiment, we recall the
connection of $\sigma_T$ and $\sigma_L$ with the proton structure
function, $F_2(W^2, Q^2)$. For $x \simeq Q^2/W^2 \ll 1$ it reads
\begin{equation}
F_2 (W^2, Q^2) \simeq {{Q^2} \over {4 \pi^2 \alpha}} 
(\sigma_T + \sigma_L) \equiv {{Q^2} \over {4 \pi^2 \alpha}} 
\sigma_{\gamma^* p}.
\end{equation} 
Following common practice, in (12) we have introduced the virtual
photoproduction cross section,
\begin{equation}
\sigma_{\gamma^* p} \equiv \sigma_T + \sigma_L.
\end{equation} 
The explicit form of $F_2(W^2, Q^2)$ is easily obtained by substituting
$\sigma_T$ and $\sigma_L$ from (7) and (10), respectively, into (12).
Its asymptotic form for $Q^2 \gg m^2_0$, inserting (9) and (11) into
(12), reads
\begin{equation}
F_2(W^2, Q^2 \gg m^2_0) = {N \over {4 \pi^2 \alpha}} 
\left( \ln {1 \over {ax}} \right)
\Biggl[ 1 + \xi \Biggl( \ln {{Q^2} \over {m^2_0}} - 1 -
{{{1\over 2} \ln^2 {{Q^2} \over {m^2_0}} + {{\pi^2} \over 6}}
\over {\ln {1 \over {ax}}}} \Biggr) \Biggr],
\end{equation} 
where the scaling variable $x \simeq Q^2/W^2$ was substituted.  The
transverse part of $F_2$, obtained by putting $\sigma_L = 0$ in (12) and
$\xi = 0$ in (14), shows scaling behavior for $Q^2$ sufficiently large,
$Q^2 \gg m^2_0$. The approach to scaling according to (7) depends on the
scale $m^2_0$ and will be less fast, if our treatment will be refined by
allowing for quark-flavor dependent threshold masses replacing the
single scale, $m_0$. We note that the strict absence of scaling
violation in the transverse part of $F_2$ for asymptotic values of $Q^2$
is related to the simplified input assumption of a $1/m^2$ decrease of
$e^+e^-$ annihilation as a function of the $e^+e^-$ energy, $\sqrt s
\equiv m^2$, as adopted in (5). If this assumption is dropped by
allowing for realistic scaling violations in $R \equiv \sigma (e^+ e^-
\to {\rm hadrons})/\sigma (e^+e^- \to \mu^+ \mu^-)$, logarithmic scaling
violations will also be induced in the transverse part of $F_2$. For the
time being, we keep our simplified form (3), (4) with (5), also in view
of the fact that experimental data separating $\sigma_T$ and $\sigma_L$
are not available so far in the region of small $x$.

For the comparison with the data from H1 and ZEUS from HERA, we have
chosen for the free parameters the values of
\begin{equation}
\begin{array}{rcl}
N &=& 5.13 \cdot 4 \pi^2\alpha = 1.48 \\[1ex]
m^2_0 &=& 0.89\,\, {\rm GeV}^2\\[1ex]
\xi &=& 0.171\\[1ex]
a &=& 15.1
\end{array}
\end{equation} 
which are obtained from fitting the H1 and ZEUS data for $F_2$
\cite{f2data} in the range of $Q^2 < 200$ GeV$^2$. The value of $m_0$,
obtained from the fit does not deviate much from the $\rho^0$ mass, as
expected from its meaning as an effective threshold of $e^+ e^-$
annihilation.  The fact that $m^2_0$ is nevertheless slightly larger
than $m_\rho^2 \simeq 0.59$ GeV$^2$ is presumably due to the
simplification of not having separated the charm contribution to
$\sigma_T$ and $\sigma_L$ in (3) and (4) with its substantially higher
threshold mass from the contribution of the light quarks. The value of
the longitudinal to transverse ratio, $\xi$, in (15) seems reasonable.

Fig.\ 2 shows remarkably good agreement of $\sigma_{\gamma^* p}$ from
(7), (10), (13), and (15) with the HERA data from H1 and ZEUS over the
full $Q^2$ range from photoproduction to $Q^2 \simeq 350$ GeV$^2$, and
energies $W$ from $W \simeq 60$ GeV to $W \simeq 245$ GeV, corresponding
to values of the scaling variable of $x \lsim 0.05$. In Fig.\ 3, we show
a comparison between Generalized Vector Dominance and the HERA data as a
function of $W$ for a series of values of $Q^2$.  Fig.\ 3 obviously
illustrates what happens if Fig.\ 2 is cut in a direction perpendicular
to the abscissa at selected values of $Q^2$.  Comparing the slope of the
theoretical predictions at fixed $W$ for different values of $Q^2$, one
observes a rising slope in this log-log plot. The rising slope
originates from the change in scale of the $W$ dependence, $a\cdot
m_0^2$ from photoproduction effectively being replaced by $a\cdot Q^2$
for $Q^2 \gg m_0^2$, in combination with the increasing importance of
$\sigma_L$.  The increase towards low $x$ becomes more dramatic when
plotting the structure function $F_2$ on a linear scale against log $x$,
compare Fig.\ 4. This Figure explicitly also shows that our theoretical
prediction for the transverse part of the structure function, $F_{2,T}
(W^2, Q^2)$, defined by ignoring $\sigma_L$ in (12), has reached its
scaling limit for $Q^2 \gsim 12$ GeV$^2$.  The rise of $F_2 (W^2, Q^2)$
with increasing $Q^2$ for $Q^2 \gsim 12$ GeV$^2$, is due to the
influence of $\sigma_L$, as discussed in connection with (14).

Refinements of the present work immediately suggest themselves. The
charm quark contribution should be treated separately in the basic
ansatz, and the low energy behavior of photoproduction and electron
scattering has to be incorporated.  Moreover, a detailed analysis of the
relation between $\sigma_{\gamma^* p}$ and diffractive production at $t
= 0$ is to be carried out, improving and elaborating upon previous
suggestions \cite{stod,schild2}.

Various parametrizations of the experimental data on low $x$ deep
inelastic scattering, including photoproduction, exist in the
literature, either based on \cite{allm,dola2} modifications of Regge
theory or on a combination \cite{badelek} of $\rho^0, \omega, \phi$
dominance with the parton-model approach. The fit of the data presented
in \cite{haidt} is of interest in the context of the present paper, as
logarithmic $Q^2$ and $x$ dependences only are employed in the fit.

In the present work we have shown that the data on low $x$ deep
inelastic scattering are consistent with a picture in which the role of
$\rho^0, \omega, \phi$ in photoproduction is extended to more massive
vector states in deep inelastic scattering, coupled to the photon with a
strength known from $e^+e^-$ annihilation.  In this sense there is
continuity in the underlying dynamics. With increasing $Q^2$, the scale
in the logarithmic $W$ dependence of photoproduction becomes gradually
replaced by $Q^2$ as soon as $Q^2$ becomes large compared with the
thresholds of light quark and charm quark production in $e^+ e^-$
annihilation. In addition, a scaling-violating longitudinal contribution
to $F_2$ is turned on. While details of this picture may have to be
refined, the principal dynamical ansatz, resting on the connection
between the virtual photon absorption cross section,
$\sigma_{\gamma^*p}$, and $e^+e^-$ annihilation into hadronic $q \bar q$
states with subsequent diffractive forward scattering on the proton, as
suggested twenty-five years ago, will be likely to stand the test of
time.

\bigskip
\noindent{\large \bf Acknowledgment}\\[1mm] 
We thank Dieter Haidt, Peter Landshoff and G\"unter Wolf for stimulating
discussions on the HERA results. We also thank the H1 and ZEUS
collaborations for providing us with their most recent $F_2$ data.  One
of us (D.S.) started to work on the subject matter of the present paper
while visiting the Max Planck Institut f{\"u}r Physik in Munich in
summer 1995.  It is a pleasure to thank Wolfhart Zimmermann for
hospitality and Leo Stodolsky for useful discussions on the subject of
this paper.

\bigskip

\newpage
\begin{figure}[htb] 
\unitlength 1mm
\begin{picture}(162,195)
\put(-1.5,0){
\epsfxsize=15.85cm
\epsfysize=18.3cm
\epsfbox{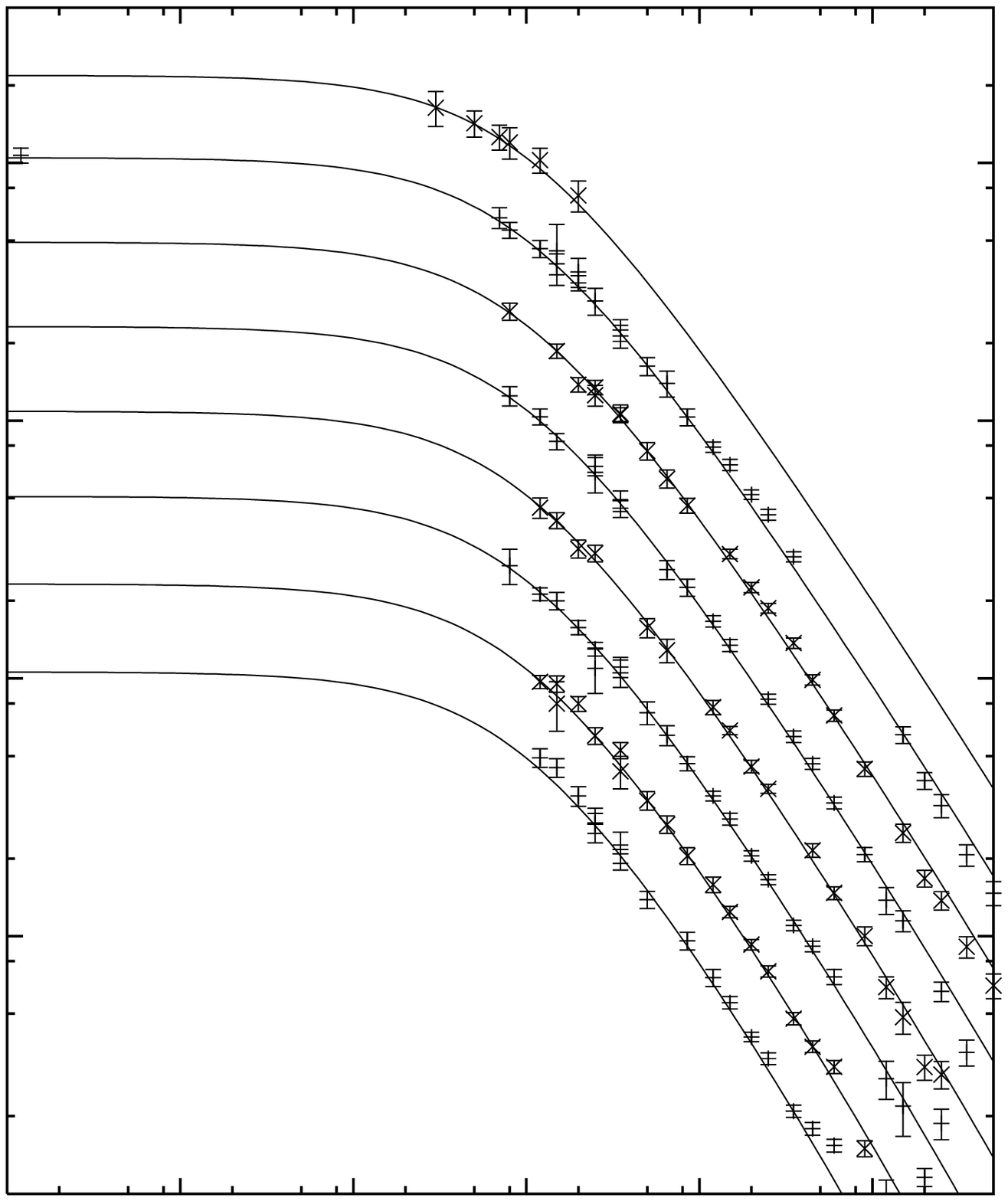}}
\end{picture}
\begin{picture}(162,195)
\put(-1.5,195.5){
\epsfxsize=15.85cm
\epsfysize=18.3cm
\epsfbox{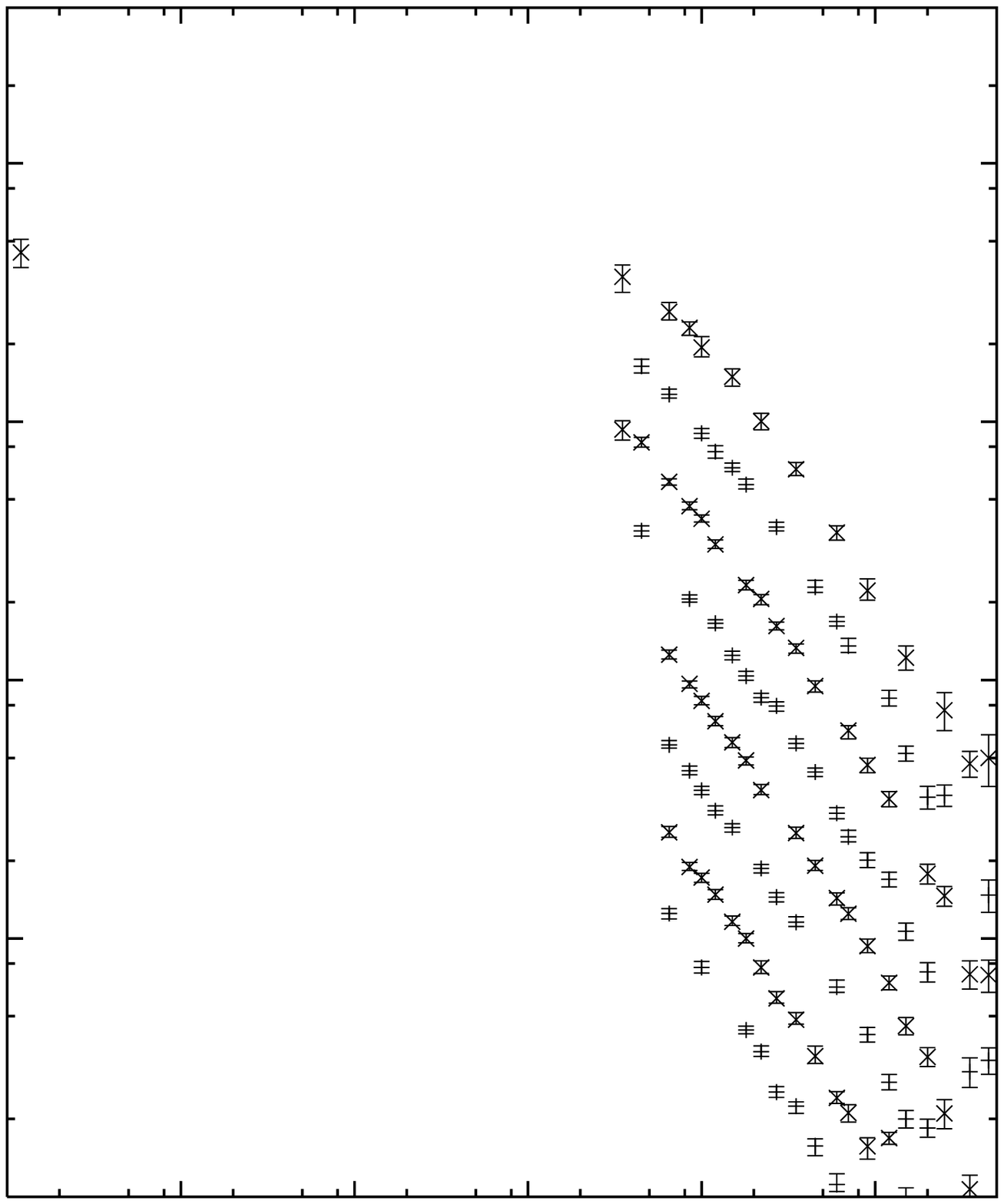}}
\put(123,199){\makebox(0,0){\large $Q^2$[GeV$^2$]}}
\put(136,204){\makebox(0,0){$10^2$}}
\put(111,204){\makebox(0,0){$10$}}
\put(87,204){\makebox(0,0){$1$}}
\put(64.5,204){\makebox(0,0){$10^{-1}$}}
\put(40,204){\makebox(0,0){$10^{-2}$}}
\put(15.5,204){\makebox(0,0){$10^{-3}$}}
\put(12,353){\makebox(0,0)[r]{$10^4$}}
\put(12,317){\makebox(0,0)[r]{$10^3$}}
\put(12,281){\makebox(0,0)[r]{$10^2$}}
\put(12,244){\makebox(0,0)[r]{$10$}}
\put(12,208){\makebox(0,0)[r]{$1$}}
\put(18,370){\makebox(0,0)[l]%
{\large $\sigma_{\gamma^*p}[\mu{\rm b}]$}}
\put(20,361){\makebox(0,0)[l]{$\times 128,~W=245$ GeV}}
\put(20,350.5){\makebox(0,0)[l]{$\times 64,~W=210$ GeV}}
\put(20,338){\makebox(0,0)[l]{$\times 32,~W=170$ GeV}}
\put(20,326){\makebox(0,0)[l]{$\times 16,~W=140$ GeV}}
\put(20,314.5){\makebox(0,0)[l]{$\times 8,~W=115$ GeV}}
\put(20,302){\makebox(0,0)[l]{$\times 4,~W=95$ GeV}}
\put(20,290){\makebox(0,0)[l]{$\times 2,~W=75$ GeV}}
\put(20,278){\makebox(0,0)[l]{$W=60$ GeV}}
\end{picture}
\vspace*{-19.35cm}
\caption{\it 
  Generalized Vector Dominance prediction for $\sigma_{\gamma^*p}$ based
  on (7), (10), (13) and (15) compared with the experimental data from
  the H1 and ZEUS collaborations at HERA.}
\label{fig2}
\end{figure}
\newpage
\begin{figure}[htb] 
\unitlength 1mm
\begin{picture}(162,195)
\put(-1.5,0){
\epsfxsize=15.85cm
\epsfysize=18.3cm
\epsfbox{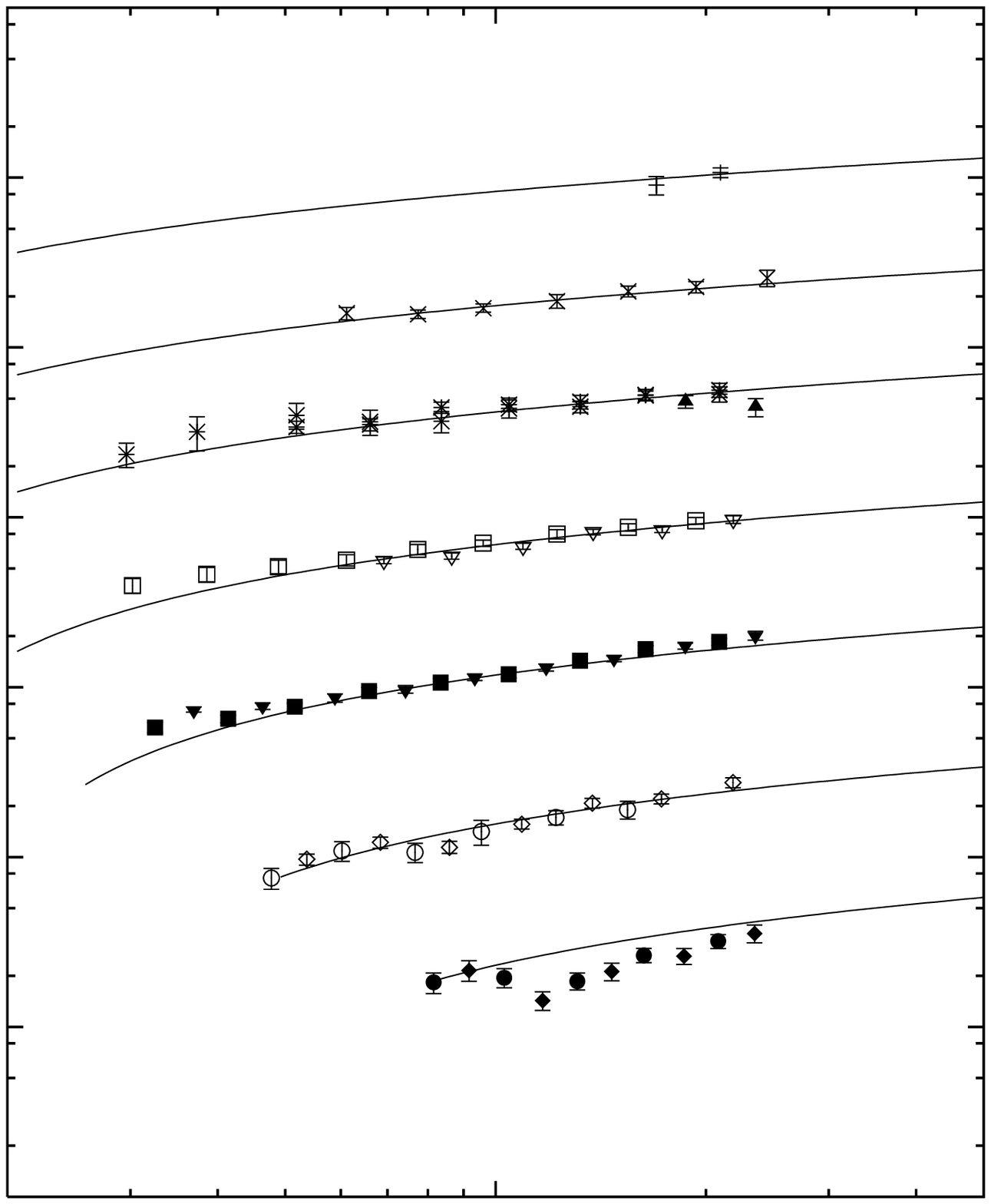}}
\setlength{\unitlength}{0.1bp}
\put(1287,1146){\makebox(0,0)[l]{$Q^2=350$ GeV${}^2$}}
\put(565,1500){\makebox(0,0)[l]{$\times 2,~~~Q^2=120$ GeV${}^2$}}
\put(565,1907){\makebox(0,0)[l]{$\times 4,~~~Q^2=35$ GeV${}^2$}}
\put(565,2478){\makebox(0,0)[l]{$\times 8,~~~Q^2=12$ GeV${}^2$}}
\put(565,3109){\makebox(0,0)[l]{$\times 16,~~Q^2=3.5$ GeV${}^2$}}
\put(565,3904){\makebox(0,0)[l]{$\times 32,~~Q^2=1.2$ GeV${}^2$}}
\put(565,4342){\makebox(0,0)[l]{$\times 64,~~Q^2=0$}}
\put(565,4934){\makebox(0,0)[l]{\large $\sigma_{\gamma^*p}[\mu {\rm b}]$}}
\put(3793,51){\makebox(0,0){\large $W$[GeV]}}
\put(435,251){\makebox(0,0){$20$}}
\put(2394,251){\makebox(0,0){$100$}}
\put(4329,251){\makebox(0,0){$500$}}
\put(400,5084){\makebox(0,0)[r]{$10^5$}}
\put(400,4408){\makebox(0,0)[r]{$10^4$}}
\put(400,3732){\makebox(0,0)[r]{$10^3$}}
\put(400,3056){\makebox(0,0)[r]{$10^2$}}
\put(400,2379){\makebox(0,0)[r]{$10$}}
\put(400,1703){\makebox(0,0)[r]{$1$}}
\put(400,1027){\makebox(0,0)[r]{$10^{-1}$}}
\put(400,351){\makebox(0,0)[r]{$10^{-2}$}}
\end{picture}
\caption{\it 
  Same as Fig.\ 2, but for various values of $Q^2$ as a function of
  $W$.}
\label{fig3}
\end{figure}

\begin{figure}[htbp] 
\unitlength 1mm
\vspace*{-2cm}
\begin{picture}(162,220)(5,0)
\put(8.5,140){
\epsfxsize=7.4cm
\epsfysize=7.1cm
\epsfbox{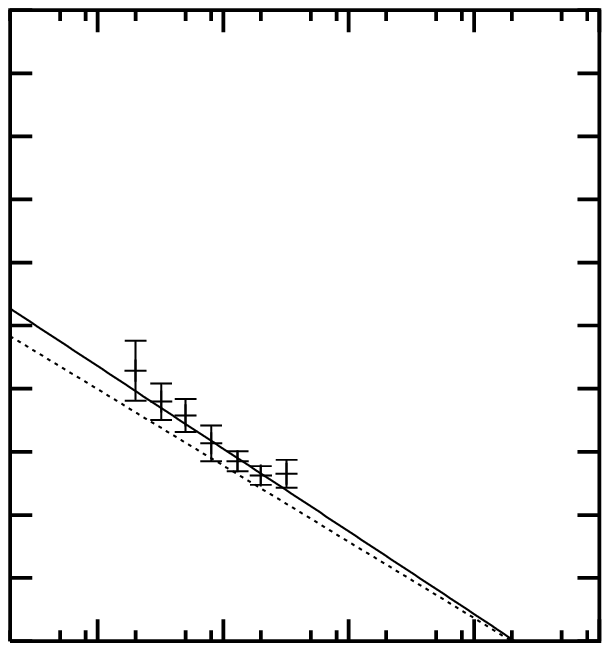}}
\put(23,202){\makebox(0,0)[l]{\large $F_2$}}
\put(40,194){\makebox(0,0)[l]{\large $Q^2 = 1.2$ GeV${}^2$}}
\put(73,144){\makebox(0,0){\large $x$}}
\put(79,147){\makebox(0,0){$10^{-1}$}}
\put(66.5,147){\makebox(0,0){$10^{-2}$}}
\put(54.5,147){\makebox(0,0){$10^{-3}$}}
\put(42,147){\makebox(0,0){$10^{-4}$}}
\put(30,147){\makebox(0,0){$10^{-5}$}}
\put(18,207.5){\makebox(0,0)[r]{2}}
\put(18,202){\makebox(0,0)[r]{1.8}}
\put(18,196.5){\makebox(0,0)[r]{1.6}}
\put(18,190.5){\makebox(0,0)[r]{1.4}}
\put(18,185){\makebox(0,0)[r]{1.2}}
\put(18,179){\makebox(0,0)[r]{1}}
\put(18,173.5){\makebox(0,0)[r]{0.8}}
\put(18,168){\makebox(0,0)[r]{0.6}}
\put(18,162.5){\makebox(0,0)[r]{0.4}}
\put(18,157){\makebox(0,0)[r]{0.2}}
\put(90,140){
\epsfxsize=7.4cm
\epsfysize=7.1cm
\epsfbox{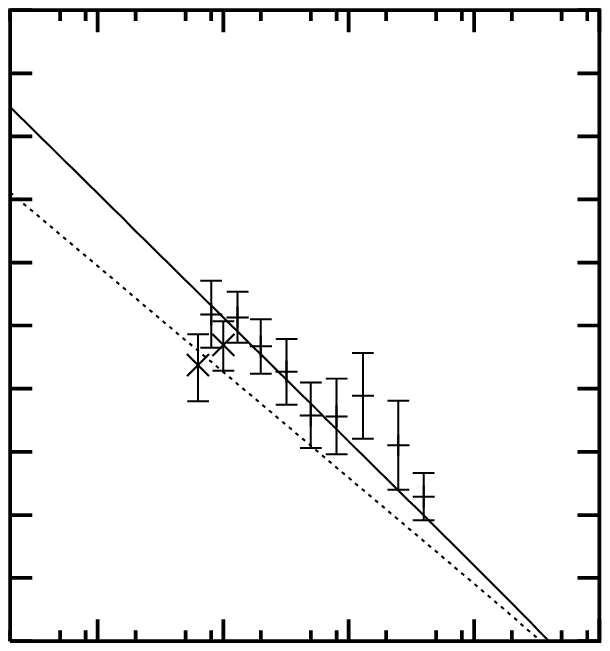}}
\put(105,202){\makebox(0,0)[l]{\large $F_2$}}
\put(120,194){\makebox(0,0)[l]{\large $Q^2 = 3.5$ GeV${}^2$}}
\put(153,144){\makebox(0,0){\large $x$}}
\put(159,147){\makebox(0,0){$10^{-1}$}}
\put(146.5,147){\makebox(0,0){$10^{-2}$}}
\put(134.5,147){\makebox(0,0){$10^{-3}$}}
\put(122,147){\makebox(0,0){$10^{-4}$}}
\put(110,147){\makebox(0,0){$10^{-5}$}}
\put(100,207.5){\makebox(0,0)[r]{2}}
\put(100,202){\makebox(0,0)[r]{1.8}}
\put(100,196.5){\makebox(0,0)[r]{1.6}}
\put(100,190.5){\makebox(0,0)[r]{1.4}}
\put(100,185){\makebox(0,0)[r]{1.2}}
\put(100,179){\makebox(0,0)[r]{1}}
\put(100,173.5){\makebox(0,0)[r]{0.8}}
\put(100,168){\makebox(0,0)[r]{0.6}}
\put(100,162.5){\makebox(0,0)[r]{0.4}}
\put(100,157){\makebox(0,0)[r]{0.2}}
\put(8.5,70){
\epsfxsize=7.4cm
\epsfysize=7.1cm
\epsfbox{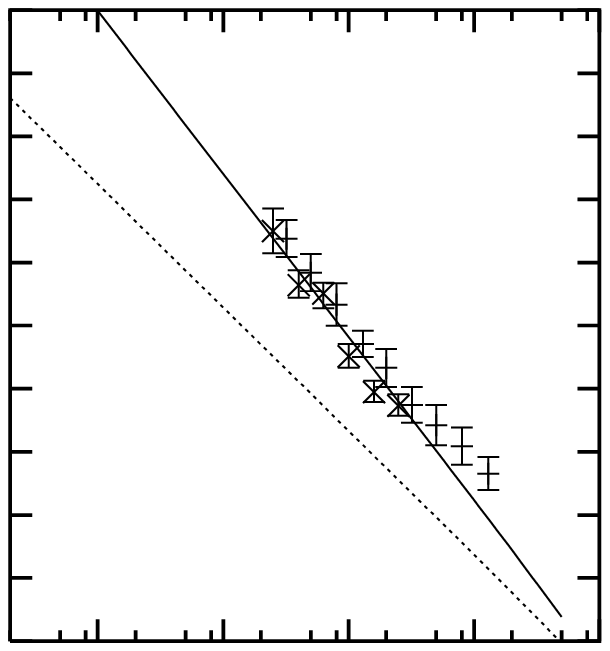}}
\put(23,132){\makebox(0,0)[l]{\large $F_2$}}
\put(26,91){\makebox(0,0)[l]{\large $Q^2 = 12$ GeV${}^2$}}
\put(73,74){\makebox(0,0){\large $x$}}
\put(79,77){\makebox(0,0){$10^{-1}$}}
\put(66.5,77){\makebox(0,0){$10^{-2}$}}
\put(54.5,77){\makebox(0,0){$10^{-3}$}}
\put(42,77){\makebox(0,0){$10^{-4}$}}
\put(30,77){\makebox(0,0){$10^{-5}$}}
\put(18,137.5){\makebox(0,0)[r]{2}}
\put(18,132){\makebox(0,0)[r]{1.8}}
\put(18,126.5){\makebox(0,0)[r]{1.6}}
\put(18,120.5){\makebox(0,0)[r]{1.4}}
\put(18,115){\makebox(0,0)[r]{1.2}}
\put(18,109){\makebox(0,0)[r]{1}}
\put(18,103.5){\makebox(0,0)[r]{0.8}}
\put(18,98){\makebox(0,0)[r]{0.6}}
\put(18,92.5){\makebox(0,0)[r]{0.4}}
\put(18,87){\makebox(0,0)[r]{0.2}}
\put(90,70){
\epsfxsize=7.4cm
\epsfysize=7.1cm
\epsfbox{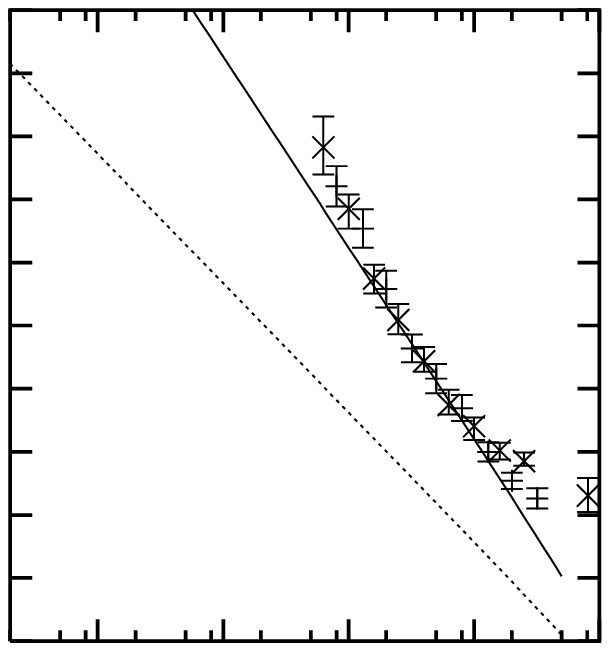}}
\put(105,132){\makebox(0,0)[l]{\large $F_2$}}
\put(108,91){\makebox(0,0)[l]{\large $Q^2 = 35$ GeV${}^2$}}
\put(153,74){\makebox(0,0){\large $x$}}
\put(159,77){\makebox(0,0){$10^{-1}$}}
\put(146.5,77){\makebox(0,0){$10^{-2}$}}
\put(134.5,77){\makebox(0,0){$10^{-3}$}}
\put(122,77){\makebox(0,0){$10^{-4}$}}
\put(110,77){\makebox(0,0){$10^{-5}$}}
\put(100,137.5){\makebox(0,0)[r]{2}}
\put(100,132){\makebox(0,0)[r]{1.8}}
\put(100,126.5){\makebox(0,0)[r]{1.6}}
\put(100,120.5){\makebox(0,0)[r]{1.4}}
\put(100,115){\makebox(0,0)[r]{1.2}}
\put(100,109){\makebox(0,0)[r]{1}}
\put(100,103.5){\makebox(0,0)[r]{0.8}}
\put(100,98){\makebox(0,0)[r]{0.6}}
\put(100,92.5){\makebox(0,0)[r]{0.4}}
\put(100,87){\makebox(0,0)[r]{0.2}}
\put(8.5,0){
\epsfxsize=7.4cm
\epsfysize=7.1cm
\epsfbox{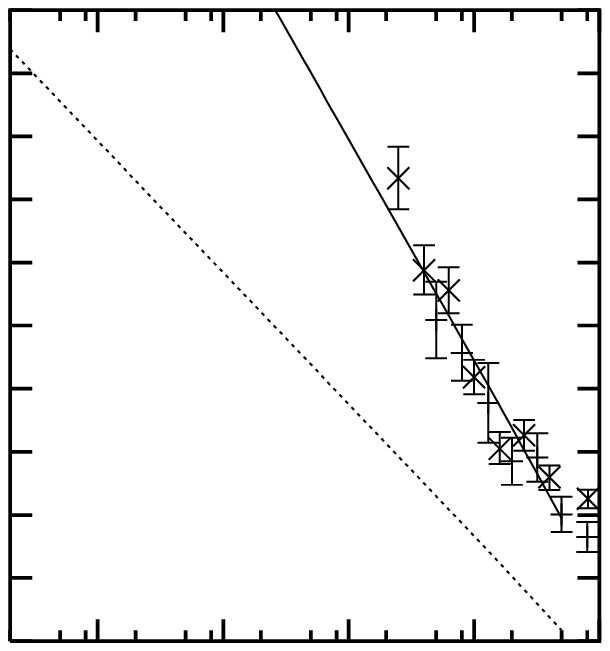}}
\put(23,62){\makebox(0,0)[l]{\large $F_2$}}
\put(26,21){\makebox(0,0)[l]{\large $Q^2 = 120$ GeV${}^2$}}
\put(73,4){\makebox(0,0){\large $x$}}
\put(79,7){\makebox(0,0){$10^{-1}$}}
\put(66.5,7){\makebox(0,0){$10^{-2}$}}
\put(54.5,7){\makebox(0,0){$10^{-3}$}}
\put(42,7){\makebox(0,0){$10^{-4}$}}
\put(30,7){\makebox(0,0){$10^{-5}$}}
\put(18,67.5){\makebox(0,0)[r]{2}}
\put(18,62){\makebox(0,0)[r]{1.8}}
\put(18,56.5){\makebox(0,0)[r]{1.6}}
\put(18,50.5){\makebox(0,0)[r]{1.4}}
\put(18,45){\makebox(0,0)[r]{1.2}}
\put(18,39){\makebox(0,0)[r]{1}}
\put(18,33.5){\makebox(0,0)[r]{0.8}}
\put(18,28){\makebox(0,0)[r]{0.6}}
\put(18,22.5){\makebox(0,0)[r]{0.4}}
\put(18,17){\makebox(0,0)[r]{0.2}}
\put(90,0){
\epsfxsize=7.4cm
\epsfysize=7.1cm
\epsfbox{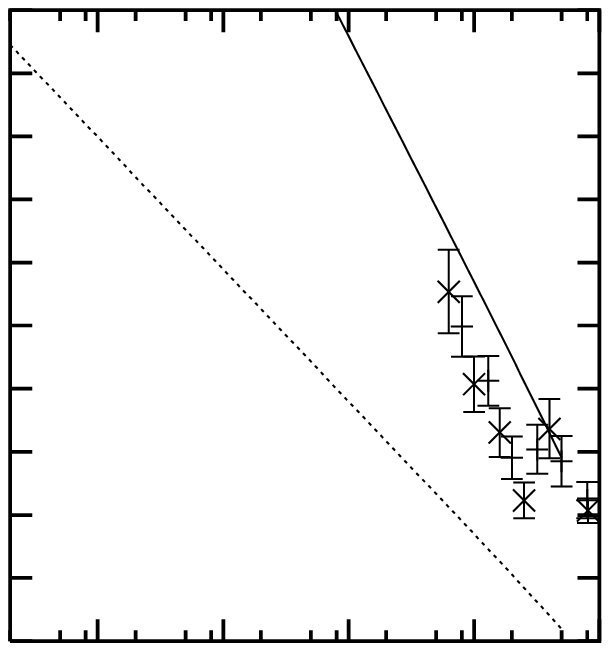}}
\put(105,62){\makebox(0,0)[l]{\large $F_2$}}
\put(108,21){\makebox(0,0)[l]{\large $Q^2 = 350$ GeV${}^2$}}
\put(153,4){\makebox(0,0){\large $x$}}
\put(159,7){\makebox(0,0){$10^{-1}$}}
\put(146.5,7){\makebox(0,0){$10^{-2}$}}
\put(134.5,7){\makebox(0,0){$10^{-3}$}}
\put(122,7){\makebox(0,0){$10^{-4}$}}
\put(110,7){\makebox(0,0){$10^{-5}$}}
\put(100,67.5){\makebox(0,0)[r]{2}}
\put(100,62){\makebox(0,0)[r]{1.8}}
\put(100,56.5){\makebox(0,0)[r]{1.6}}
\put(100,50.5){\makebox(0,0)[r]{1.4}}
\put(100,45){\makebox(0,0)[r]{1.2}}
\put(100,39){\makebox(0,0)[r]{1}}
\put(100,33.5){\makebox(0,0)[r]{0.8}}
\put(100,28){\makebox(0,0)[r]{0.6}}
\put(100,22.5){\makebox(0,0)[r]{0.4}}
\put(100,17){\makebox(0,0)[r]{0.2}}
\end{picture}
\caption{\it
  Generalized Vector Dominance prediction for $F_2$ as a function of
  $x$ compared with the HERA data for different values of $Q^2$. The
  dotted line shows the contribution to $F_2$ due to transverse virtual
  photons $(\xi = 0)$.}
\label{fig4}
\end{figure}

\end{document}